\documentclass[prd,nofootinbib,preprint,nofootinbib,superscriptaddress,nobalancelastpage,showpacs,showkeys]{revtex4}
\pdfoutput=1

\usepackage{amsmath}
\usepackage{amssymb}
\usepackage{graphicx}
\usepackage{multirow}
\usepackage{longtable}
\usepackage{subfigure}
\usepackage{bbm}
\usepackage{mathrsfs}
\usepackage{color}
\usepackage{hyperref}

\newcommand{\beq}{\begin{equation}}
\newcommand{\eeq}{\end{equation}}
\newcommand{\be}{\begin{eqnarray}}
\newcommand{\ee}{\end{eqnarray}}

\newcommand{\sss}{\sin^2 \theta_{12}}
\newcommand{\sch}{\sin^2 \theta_{13}}
\newcommand{\stch}{\sin^2 2\theta_{13}}
\newcommand{\sa}{\sin^2 \theta_{23}}

\newcommand{\ssst}{\sin^2 \theta_{12}{\mbox {(true)}}}
\newcommand{\scht}{\sin^2 \theta_{13}{\mbox {(true)}}}
\newcommand{\stcht}{\sin^2 2\theta_{13}{\mbox {(true)}}}
\newcommand{\sat}{\sin^2 \theta_{23}{\mbox {(true)}}}

\newcommand{\dcpt}{\delta_{CP}{\mbox {(true)}}}
\newcommand{\dcp}{\delta_{\mathrm{CP}}}

\newcommand{\tet}{\theta_{13}}

\def\nue{{\nu_e}}

\def\numu{{\nu_{\mu}}}
\def\anumu{{\bar\nu_{\mu}}}

\newcommand{\capdef}{}
\newcommand{\mycaption}[2][\capdef]{\renewcommand{\capdef}{#2}
       \caption[#1]{{\footnotesize #2}}}
\makeatletter
\renewcommand{\fnum@table}{\textbf{\tablename~\thetable}}
\renewcommand{\fnum@figure}{\textbf{\figurename~\thefigure}}
\makeatother

\begin{document}
\pagestyle{plain}

\preprint{IFIC-12-29}
\preprint{EURONU-WP6-12-49}

\title{Probing the Neutrino Mass Hierarchy with Super-Kamiokande}

\author{Sanjib Kumar Agarwalla}
\email{Sanjib.Agarwalla@ific.uv.es}

\author{Pilar Hern\'andez}
\email{pilar@ific.uv.es}

\affiliation{
Instituto de F\'{\i}sica Corpuscular, CSIC-Universitat de Val\`encia\\
		Apartado de Correos 22085, E-46071 Valencia, Spain}
		
\date{\today}		
		
\begin{abstract}

We show that for recently discovered large values of $\theta_{13}$, a superbeam with an average neutrino energy  of  $\sim$ 5 GeV, such as those being proposed 
at CERN, if pointing to Super-Kamiokande ($L \simeq 8770$~km), could reveal the neutrino mass hierarchy at $5\,\sigma$ in less than two years irrespective of the 
true hierarchy and CP phase. The measurement  relies on the near resonant matter effect in the $\nu_\mu \rightarrow \nu_e$ oscillation channel, and can be done 
counting the total number of appearance events with just a neutrino beam.

\end{abstract}

\pacs{14.60.Pq,14.60.Lm,13.15.+g}

\keywords{Neutrino, Superbeam, Mass Hierarchy, Super-Kamiokande, Matter Effect}


\maketitle

\section{Introduction}
\label{sec:intro}

The recent determination of $\theta_{13}$ by T2K~\cite{Abe:2011sj}, Daya Bay~\cite{An:2012eh} and RENO~\cite{Ahn:2012nd} indicates a value very close 
to the previous Chooz bound~\cite{Apollonio:1999ae}. The results of Daya Bay/RENO 
\begin{eqnarray}
\left.\sin^2 2 \theta_{13}\right|_{\rm Day Bay} &=& 0.092(16)(5),\nonumber\\
\left.\sin^2 2 \theta_{13}\right|_{\rm RENO} &=&  0.113(13)(19) ,
\end{eqnarray}
are in perfect agreement. These new results have now been incorporated in global fits of all available neutrino oscillation 
data~\cite{Tortola:2012te,Fogli:2012ua}, which indicate towards a non-zero value of $\tet$ at around $8\,\sigma$ C.L. and hint that 
$\sch$ lies between 0.019 and 0.033 at $2\,\sigma$ C.L. with a best-fit value of 0.026 for normal mass ordering~\cite{Tortola:2012te}.   
Such large value of the 1--3 mixing angle opens up the possibility to discover leptonic CP violation and determine the neutrino mass hierarchy (MH), 
normal (NH) if $\Delta m^2_{31} \equiv m^2_3 -m^2_1 >0$, or inverted (IH) if $\Delta m^2 _{31} <0$, with significantly less effort than previously thought. 

Many studies have been carried out in the last decade to establish the best strategy to perform these measurements \cite{iss}. CP violation must be searched for 
by comparing neutrino and anti-neutrino appearance oscillation probabilities in the atmospheric  range ($E/L \sim |\Delta m^2_{31}|$). This can be realistically achieved
using neutrino beams in the $\mathrm{GeV}$ range produced in accelerators.  On the other hand the MH determination is a discrete measurement that relies on the MSW 
resonance effect \cite{Wolfenstein:1977ue, MSW}, which is relevant in several accessible neutrino beams:  atmospheric neutrinos \cite{Akhmedov:1998xq}, supernova 
neutrinos \cite{Kuo:1987qu}, and also accelerator neutrino beams for sufficiently long baselines \cite{Ayres:2004js}. Cosmology can also weigh neutrinos with precision 
and future CMB and LSS cosmological measurements  have a chance to determine the light neutrino spectrum\cite{Lesgourgues:2004ps,Jimenez:2010ev}. 
Finally, the observation of  neutrinoless double beta decay with the next generation of experiments would not only imply that neutrinos are Majorana, but also that the 
hierarchy is inverted \cite{Bilenky:1996cb}.

The possibility to measure the hierarchy with atmospheric neutrinos in the proposed iron calorimeter detector at INO \cite{INO}, combined with the results from 
accelerator beam experiments T2K and NOvA, has been revised recently in light of the large value of $\theta_{13}$~\cite{Blennow:2012gj}.  
A 3$\sigma$ C.L. determination of the hierarchy is possible irrespective of the value of the CP phase, only for the most optimistic assumptions about a 
$100\,\mathrm{kt}$ INO detector and a running time of $\sim$ 10 years. 

Out of all these possibilities, the method that is probably less affected by systematic errors, or correlations with other unknown parameters, is the 
one that uses a  neutrino beam produced in an accelerator. In particular, given the large value of $\theta_{13}$, the use of more intense conventional beams, 
the so-called superbeams, would suffice. The optimization of energy and baseline for the measurement of CP violation and the determination of the hierarchy 
are somewhat in conflict. The compromise often implies some destructive interference of both measurements: not knowing the hierarchy influences the measurement 
of the CP phase and viceversa~\cite{Minakata:2001qm}.

 A very good precision on the CP violating phase can be achieved by measuring the neutrino and anti-neutrino oscillation probabilities 
 at energies around $1$ $\mathrm{GeV}$ and at relatively short baselines of  $L\sim 300$ km in a megaton-size water Cerenkov detector \cite{Coloma:2012wq}, 
 such as proposed in the  T2HK experiment  \cite{Abe:2011ts}. For these baselines matter effects are small and therefore 
 the hierarchy cannot be determined, which affects the precision achievable in the determination of the CP phase. The hierarchy on the other hand becomes 
 almost a digital measurement at the point of maximal resonant conversion, which implies higher energies and baselines of $E\sim 6$ $\mathrm{GeV}$ and 
 $L \lesssim 10^4~$km. In this case the measurement can be done with just a neutrino beam. 
  
Instead of trying to compromise both measurements, one could devise an optimal experiment for each. In this paper, we show that 
a setup that can easily determine the hierarchy could be achieved by focussing one of the proposed superbeams 
at CERN or Fermilab, towards the {\it existing} and well-tested Super-Kamiokande detector. 
This could arguably be an efficient way to determine the hierarchy, if NOvA misses it. However one should keep in mind that the baseline CERN-Kamioka is 
$8770$~km and this requires a dip angle of $\sim 42^\circ$, which could be a technical challenge. 
Such very long baselines have been considered before together with significant upgrades in the detector technologies, in the context of neutrino 
factories~\cite{Huber:2003ak,Agarwalla:2010hk}, beta-beams~\cite{Agarwalla:2006vf,Agarwalla:2007ai} and superbeams~\cite{DeJongh:2002dv}. 
The most important result of this work is to point out that an existing detector as Super-Kamiokande could do the job.

\section{Optimal $E$ and $L$ for hierarchy determination}
\label{sec:opt}

\begin{figure}[tp]
\includegraphics[width=0.6\textwidth]{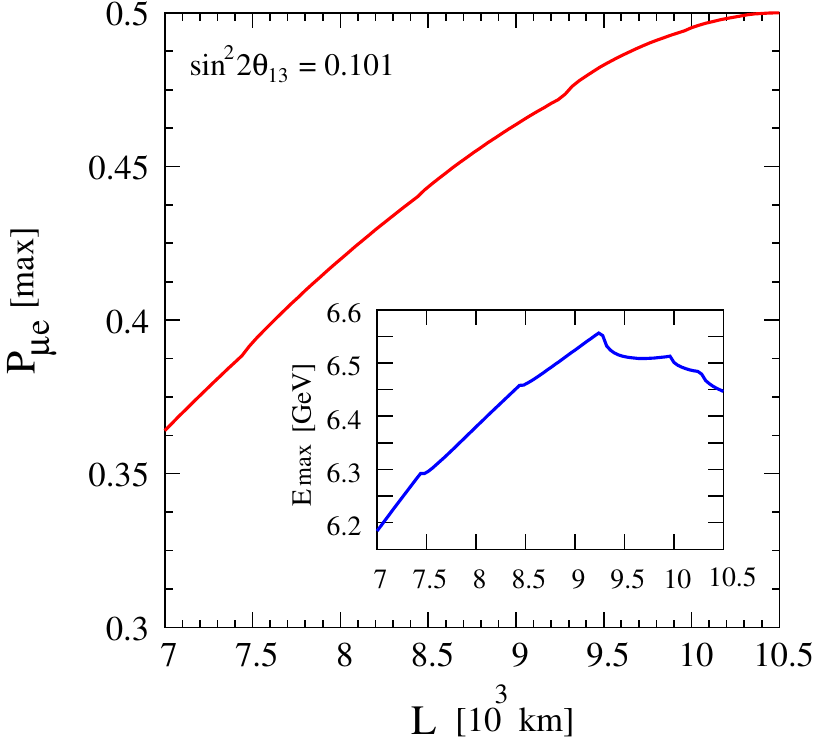}
\mycaption{\label{fig:probmax} 
In the inset, neutrino energy where $P_{\mu e}$ is maximum as a function of $L$ for NH. At the outset, the  corresponding maximum oscillation 
probability versus $L$. The results are obtained using the approximation of eq.~(\ref{eq:prob}), {\it i.e.} neglecting the solar splitting. }
\end{figure}

As is well-known, the sensitivity of neutrino oscillation probabilities to the neutrino MH relies mostly on the matter effects in neutrino 
propagation\footnote{For a similar discussion see \cite{Banuls:2001zn, PhysRevLett.94.051801}.}. Neglecting the solar mass splitting, 
the $\nu_\mu\rightarrow \nu_e$ oscillation probability for a neutrino of energy $E$ that crosses the Earth with a baseline $L$ is given by 
\begin{eqnarray}
P_{\mu e} = \sin^2\theta_{23}  \sin^2 2 \tilde{\theta}_{13} \sin^2 \left({\Delta \tilde{m}_{31}^2 L\over 4 E}\right)   
\label{eq:prob}
\end{eqnarray}
using constant line-averaged Earth matter density, with
\begin{eqnarray}
\Delta \tilde{m}_{31}^2 &\equiv& \sqrt{(\Delta m^2_{31} \cos 2\theta_{13} -A)^2+ (\Delta m^2_{31} \sin 2 \theta_{13})^2} ,\nonumber\\
\sin^2 2 \tilde{\theta}_{13} &\equiv& \sin^2 2 \theta_{13} \left({\Delta m^2_{31} \over \Delta \tilde{m}_{31}^2}\right)^2 
\end{eqnarray}
and $A \equiv 2 \sqrt{2} G_F n_e(L) E$. $G_F$ is the Fermi constant and $n_e$ is the electron number density in the Earth which depends on $L$.  
Let us first consider the NH, that is $\Delta m^2_{31} >0$. In order to maximize the neutrino oscillation probability, it is not only necessary to choose 
the resonance energy which maximizes the effective angle in matter 
\begin{eqnarray}
\sin^2 2 {\tilde \theta}_{13}|_{E=E_{\rm res}} = 1, \;\;\; E_{\rm res} \equiv {  \Delta m^2_{31} \cos 2\theta_{13} \over 2 \sqrt{2} G_F n_e},
\end{eqnarray}
but also we need to maximize the oscillatory term by choosing $L=L_{max}$ \cite{Banuls:2001zn}:
\begin{eqnarray}
\left. n_e(L) L\right|_{L_{\rm max}} = {~\pi\over \sqrt{2}G_F \tan 2 \theta_{13}}. 
\end{eqnarray}
Using the best-fit value of $\stch = 0.101$ from~\cite{Tortola:2012te}, we find $L_{\rm max} \sim 10^4$~km and $E_{\rm res}\sim 6.6$ $\mathrm{GeV}$. 
It is very important to note that the maximization of the probability would not have been achieved in terrestrial distances had $\tet$ turned out 
to be smaller. 

If we impose such resonance conditions, the oscillation probability for the IH, $\Delta m^2_{31} < 0$, would be very much suppressed since
\begin{eqnarray}
{\rm IH}: \sin^2 2 {\tilde \theta}_{13}|_{E=E_{\rm res}} \simeq {\sin^2 2 \theta_{13}\over  1+ 3\cos^2 2 \theta_{13}}.
\end{eqnarray}
Since also for the IH case, the oscillatory term would be at most 1, the oscillation probability would be widely different for NH and IH:
\begin{eqnarray}
P_{\mu e} ({\rm NH})|_{E_{\rm res}, L_{\rm max}} &\simeq& {1 \over 2}, \nonumber\\
P_{\mu e} ({\rm IH})|_{E_{\rm res}, L_{\rm max}} &\leq& {\sin^2 2\theta_{13} \over 2 (1+ 3 \cos^2 2 \theta_{13})} \simeq 0.014.
\end{eqnarray}
This analysis shows that if we have a neutrino beam with $E =E_{\rm res}$ and $L=L_{\rm max}$, the measurement is just a digital measurement. 
Of course in real life choosing $L_{\rm max}$ and $E_{\rm res}$ is difficult.

The existing baselines such as CERN-Kamioka ($L=8770$ km) and Fermilab-Kamioka ($L=9160$ km) lie somewhat below $L_{\rm max}$.  
In the inset of Fig.~\ref{fig:probmax}, we show the neutrino energy for which the $\numu \to \nue$ oscillation probability (as given by 
eq.~(\ref{eq:prob})) is maximum for different choices of baseline. Here we consider NH. At the outset, we depict the corresponding maximum 
oscillation probability as a function of $L$. In going from $L_{\rm max}$ to the CERN-Kamioka baseline, we see that the probability is reduced 
by a few per cent from its maximal value, and the optimal energy remains almost same.

An additional advantage of decreasing the baseline is that we get closer to the so-called magic baseline \cite{Huber:2003ak}. 
This is the baseline where solar oscillatory terms are strongly suppressed (and therefore the dependence on the CP phase):
\begin{eqnarray}
\left. n_e(L) L\right|_{L_{\rm magic}} = {  \sqrt{2}\pi \over  G_F } , ~ÊL_{\rm magic} \simeq 7690~{\rm km},
\end{eqnarray}
and this choice of magic baseline is independent of the neutrino energy. Note that, we have neglected the solar splitting in the previous analysis. 
This is quite a good approximation for the large value of $\theta_{13}$ that we consider and because of the fact that we are close to this magic baseline.

The existing baselines CERN-Kamioka and Fermilab-Kamioka happen to be in the right ballpark, where the matter enhancement is still large while 
being close to the magic baseline, the sensitivity to the hierarchy will be all most independent of the CP phase. 
It makes therefore all the sense to ask the question how well  the existing Super-Kamiokande can resolve the neutrino MH
in conjunction with long baseline superbeam. 

\begin{table}[tp]
\begin{center}
\begin{tabular}{||c||c||} \hline\hline
\multicolumn{1}{||c||}{{\rule[0mm]{0mm}{6mm}{Central (true) Values}}}
& \multicolumn{1}{c||}{\rule[-3mm]{0mm}{6mm}{External $1\,\sigma$ error}}
\cr
\hline \hline
$\scht = 0.026$ & $\sigma(\sch)=13\%$ \cr 
\hline
$\Delta m^2_{31}{\rm (true)} = 2.53 \times 10^{-3} \ {\rm eV}^2$ (NH) & $\sigma(\Delta m^2_{31})= 4\%$ \cr
\hline
$\Delta m^2_{31}{\rm (true)} = -2.40 \times 10^{-3} \ {\rm eV}^2$ (IH) & $\sigma(\Delta m^2_{31})= 4\%$ \cr
\hline
$\sat = 0.49$ (NH) & $\sigma(\sa)=8\%$ \cr
\hline
$\sat = 0.53$ (IH) & $\sigma(\sa)=8\%$ \cr
\hline
$\Delta m^2_{21}{\rm (true)} = 7.62 \times 10^{-5} \ {\rm eV}^2$ & $\sigma(\Delta m^2_{21})=3\%$ \cr
\hline
$\ssst= 0.32$ & $\sigma(\sss)=5\%$ \cr
\hline
$\rho{\rm (true)} = 1$ & $\sigma(\rho)=5\%$ \cr
\hline \hline
\end{tabular}
\mycaption{\label{tab:bench}
Best-fit values of oscillation parameters \& their $1\,\sigma$ estimated errors. 
In the last row, $\rho$ is the Earth matter density, relative to the value given by the PREM profile.}
\end{center}
\end{table}

For all the simulation results presented in this paper, we calculate the exact three generation oscillation probability using the full
realistic PREM~\cite{Dziewonski:1981xy} profile for the Earth matter density. Unless stated otherwise, for all calculations, we use 
the central (true) values of the oscillation parameters given in the first column of Table~\ref{tab:bench} which are obtained 
from the global fit of world neutrino data~\cite{Tortola:2012te}. We vary $\dcpt$ in its allowed range of 0-2$\pi$ and study both the 
mass hierarchies. In all fits, we marginalize over {\it all} oscillation parameters, the Earth matter density by allowing all of these 
to vary freely in the fit. We impose simple Gaussian priors on these parameters, with the corresponding $1\,\sigma$ errors as
mentioned in the second column of Table~\ref{tab:bench} which are taken from~\cite{Tortola:2012te}. 
Note that, in our study, we have imposed a prior on $\scht$ with the $1\,\sigma$ error of 13\% based on the information 
from~\cite{Tortola:2012te}, but its impact is  marginal. The external information on the Earth matter density ($\rho$) is
assumed to come from the study of the tomography of the Earth~\cite{Geller:2001ix,Ohlsson:2003ip}. In the fit, we allow
for a 5\% uncertainty in the PREM profile and take it into account by inserting a prior and marginalizing over the density normalization. 
The CP phase $\dcp$ is completely free in the marginalization.
 
\begin{figure}[tp]
\includegraphics[width=5.4cm, height=5.75cm]{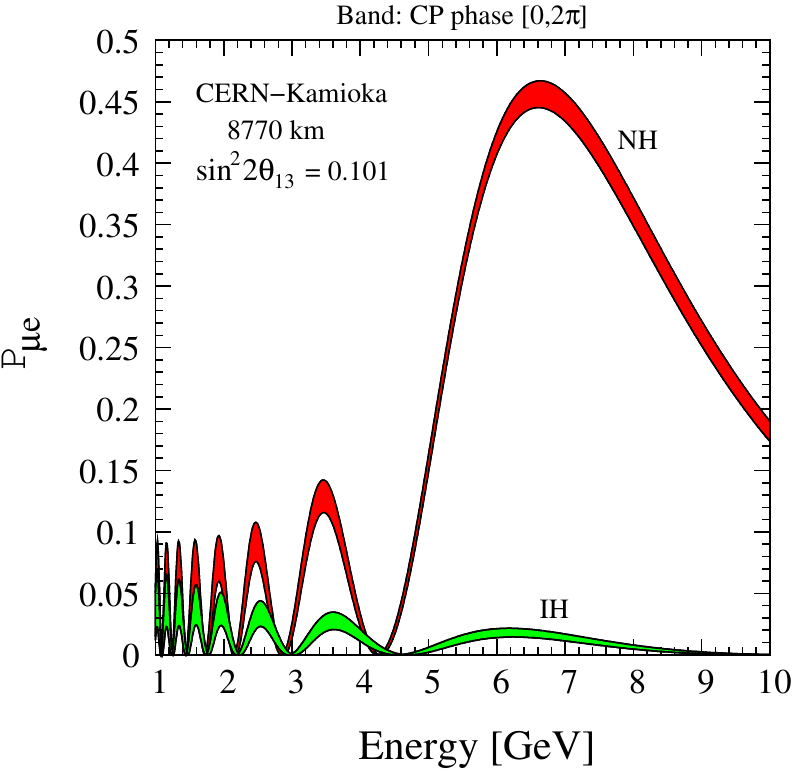}
\includegraphics[width=5.4cm, height=5.75cm]{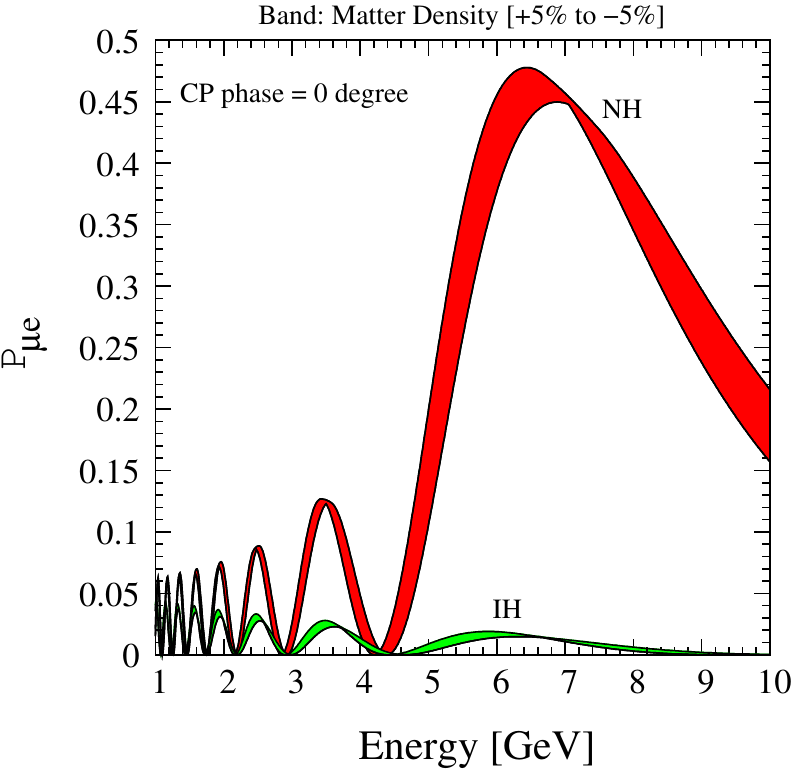}
\includegraphics[width=5.4cm, height=5.75cm]{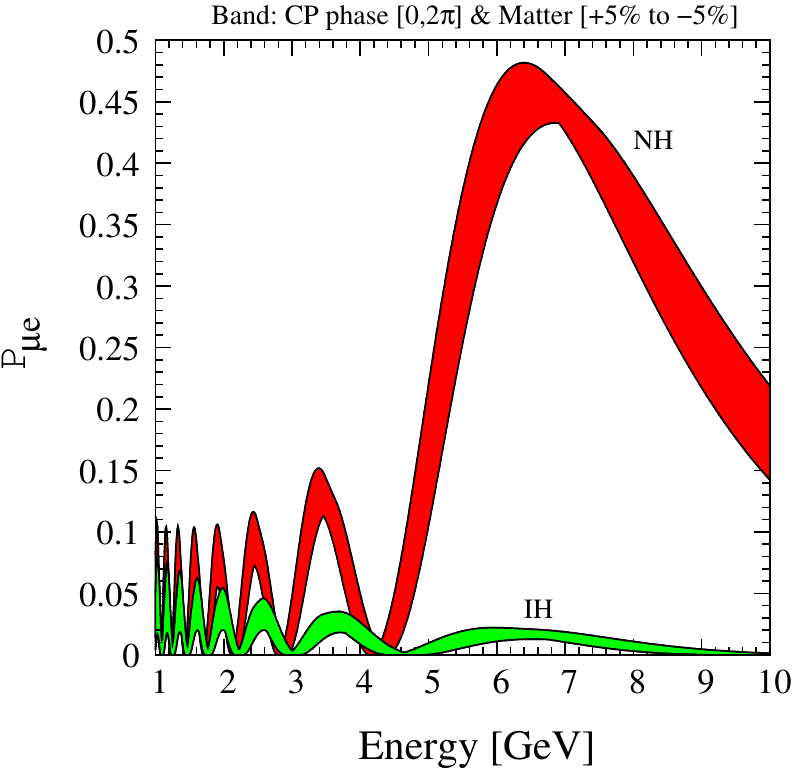}
\mycaption{\label{fig:probability} $P_{\mu e}$ as a function of $E$ for the CERN-Kamioka baseline.
We have taken $\stch=0.101$ and for all other oscillation parameters we assume the central values of Table~\ref{tab:bench}.
In left panel, the band portrays the effect of unknown $\dcp$. In middle panel, the band shows the effect of $\pm$ 5\% uncertainty in the PREM 
profile on $P_{\mu e}$. The combined effect of unknown $\dcp$ and $\pm$ 5\% uncertainty in the PREM profile is depicted as a band in right panel.}
\end{figure}

In Fig.~\ref{fig:probability}, we show the full three-flavor oscillation probability $\numu \rightarrow \nue$ using the PREM~\cite{Dziewonski:1981xy}
density profile for the CERN-Kamioka baseline as a function of neutrino energy. We allow $\dcp$ to vary in its entire range of $0$ to $2\pi$ and the resultant
probability is shown as a band in left panel of Fig.~\ref{fig:probability}, with the thickness of the band reflecting the effect of $\dcp$ on $P_{\mu e}$. 
Since this baseline is close to the magic baseline, the effect of the CP phase is seen to be almost negligible. This figure is drawn assuming the benchmark 
values of the oscillation parameters given in Table~\ref{tab:bench}. We present the probability for both NH and IH. As expected the probability for the NH is a bit lower than 
$1/2$ but still close to this maximal value. The probability for NH is hugely enhanced for the neutrinos, while for the IH matter effects do not bring
any significant change which is the key to distinguish between NH and IH. In middle panel of Fig.~\ref{fig:probability}, the band shows the effect of 
$\pm$ 5\% uncertainty in the PREM profile on $P_{\mu e}$. The combined effect of unknown $\dcp$ and $\pm$ 5\% uncertainty in the PREM profile 
on $P_{\mu e}$ is depicted as a band in right panel of Fig.~\ref{fig:probability}.

\section{Experimental Setup}
\label{sec:setup}

\subsection{CERN based Superbeam}
\label{sec:superbeam}

Superbeams are conventional neutrino and anti-neutrino beams created from the decay of horn-focused pions.
This superbeam technique is well established and understood. Future possibilities of having high-intensity proton sources at CERN 
and their relevance for high intensity neutrino beams were discussed in detail in Ref.~\cite{Rubbia:2010fm}.
In the context of the planned LHC upgrades, the CERN accelerator infrastructures hopefully will be able
to supply high power beams. One of the possibilities discussed in \cite{Rubbia:2010fm} is a new high power accelerator 
(HP-PS2), with a proton energy of $50\,\mathrm{GeV}$ and a beam power of $1.6\,\mathrm{MW}$, resulting in $3\times10^{21}$ 
protons-on-target (pot) per year. A conceptual design is being developed in the context of the LAGUNA-LBNO design 
study~\cite{Angus:2010sz,Rubbia:2010zz}. We will take this design as our reference setup, although obviously there are other alternatives. 
The expected neutrino fluxes from this machine have been estimated for the CERN-Pyh{\"a}salmi baseline by A. Longhin 
in~\cite{Longhin:2010zz,longhin}. We have simply scaled these fluxes to the longer $L=8770$ km baseline, as $L^{-2}$. 
There is however room for beam optimization, specially in what regards the neutrino energy, which could probably be increased. 
We will take as our reference an integrated luminosity of $5 \times 10^{21}$ pot. 

\subsection{The Super-Kamiokande detector}
\label{sec:sk}

Here in our study, we consider the existing and well understood Super-Kamiokande detector~\cite{Fukuda:1998mi,Ashie:2005ik} 
with a fiducial mass of $22.5\,\mathrm{kt}$. In this work, we only rely on total signal and background event rates and no spectral 
information has been used. To estimate the total charged current (CC) signal event rates due to $\nue$ appearance in the true neutrino energy window of 
$0.5\,\mathrm{GeV}$ to $10\,\mathrm{GeV}$, we use the pre-cut efficiencies as given in Table~1 of~\cite{Dufour:2010vr}, in the context of 
the T2KK proposal. These efficiencies are estimated based on the criteria that the events are fully contained inside the fiducial volume, have a 
single Cerenkov ring recognized as electron-like, and with no Michel electron present. It gives an average efficiency of 37\%
in the $5\,\mathrm{GeV}$ to $10\,\mathrm{GeV}$ true neutrino energy window where we have the maximum near resonant matter effect.
For $\nue$ appearance channel, we have considered three different types of background: intrinsic $\nue$ contamination of the beam (Int) 
with the same efficiency as of signal, the number of muon events which will be misidentified as electron events (Mis-id), and neutral current (NC)
events. In order to calculate the effective number of Mis-id and NC backgrounds in the true neutrino energy window of 
$0.5\,\mathrm{GeV}$ to $10\,\mathrm{GeV}$, we take the corresponding efficiencies from Table~1 of~\cite{Dufour:2010vr}.
Just to give an idea, in $5\,\mathrm{GeV}$ to $10\,\mathrm{GeV}$ true neutrino energy window, the NC background rejection efficiency is 90\%. 

In~\cite{Dufour:2010vr}, further cuts are applied to improve the signal-to-noise ratio for the appearance signal. 
These however have been optimized for the lower-energy JHF beam, so it would be necessary to adapt them for the neutrino beam considered 
in this work. For example, one could easily think of further improvements in the reduction of the dominant NC background
from kinematical cuts ({\it e.g.} setting lower and upper threshold of reconstructed energy). The use of multi-ring events would also most probably 
provide valuable information. Since such optimizations require a detailed Monte Carlo study of the Super-Kamiokande detector, 
we take the conservative approach of using only the pre-cut efficiencies from~\cite{Dufour:2010vr} and leave the possibility of using 
optimized kinematical cuts for a more detailed study. 

We also include the information coming from the $\numu$ disappearance channel with the same signal efficiency that 
we have considered for electrons. For this channel, NC events are the only source of background because the 
$\anumu$ `wrong-sign' contamination is negligible. The NC background to the muon signal is expected to be significantly 
smaller because the single-$\pi^{0}$ contamination is negligible in this case. We have assumed 10 times more suppression 
in the NC background compared to the $\nue$ appearance case. In any case, the disappearance signal gives a very small 
contribution towards the total MH sensitivity.

\section{Results}
\label{sec:results}

\subsection{Event rates for CERN-Kamioka baseline}
\label{sec:event}

\begin{table}[tp]
\begin{center}
\begin{tabular}{|c|c|c|} \hline\hline
\multirow{3}{*}{Channel} & \multicolumn{2}{c|}{{\rule[0mm]{0mm}{6mm}CERN-Kamioka ($8870\,\mathrm{km}$)}}\cr
\cline{2-3}
& Signal & Background\cr
\cline{2-3}
& CC-1 ring & Int+Mis-id+NC = Total\cr
\hline\hline
$\numu \to \nue$ (NH) & {\bf 40} & 1+2+16={\bf 19} \cr
\hline
$\numu \to \nue$ (IH) & {\bf 2} & 1+3+16={\bf 20} \cr
\hline\hline
$\numu \to \numu$ (NH) & {\bf 84} & {\bf 2} \cr
\hline
$\numu \to \numu$ (IH) & {\bf 89} & {\bf 2} \cr
\hline\hline
\end{tabular}
\mycaption{\label{tab:event}
First two rows show the total signal and background event rates in the $\nue$ appearance channel taking $\stch$ = 0.101 and $\dcp$ = 0$^{\circ}$.
Here `Int' means intrinsic beam contamination, `Mis-id' means misidentified muon events and `NC' stands for neutral current.
The last two rows depict the same for $\numu$ disappearance channel with only NC background.
Results are shown for both the choices of the mass ordering with total exposure of $5\times10^{21}$ pot.}
\end{center}
\end{table}

In Table~\ref{tab:event}, we show the expected number of signal and background event rates for appearance and disappearance 
channels for both the choices of the hierarchies using the pre-cut efficiencies of the Super-Kamiokande detector as described in
Sec.~\ref{sec:sk}. To estimate these event rates, we consider $\stch$ = 0.101 and $\dcp$ = 0$^{\circ}$.
For all other oscillation parameters, we have used the benchmark true values as given in Table~\ref{tab:bench}. 
It is quite evident from Table~\ref{tab:event} that there is a huge difference between the expected neutrino signal event rates for 
NH and IH in the appearance channel due to the presence of near resonant matter effect in the case of neutrino with NH. 
As far as the background is concerned, the main background contribution comes from NC events while the background coming from 
other two sources are quite small as can be seen from Table~\ref{tab:event}. For disappearance channel, the difference between the
signal event rates for NH and IH is not very large leading to a very small contribution towards the MH sensitivity.

\begin{figure}[tp]
\includegraphics[width=0.6\textwidth]{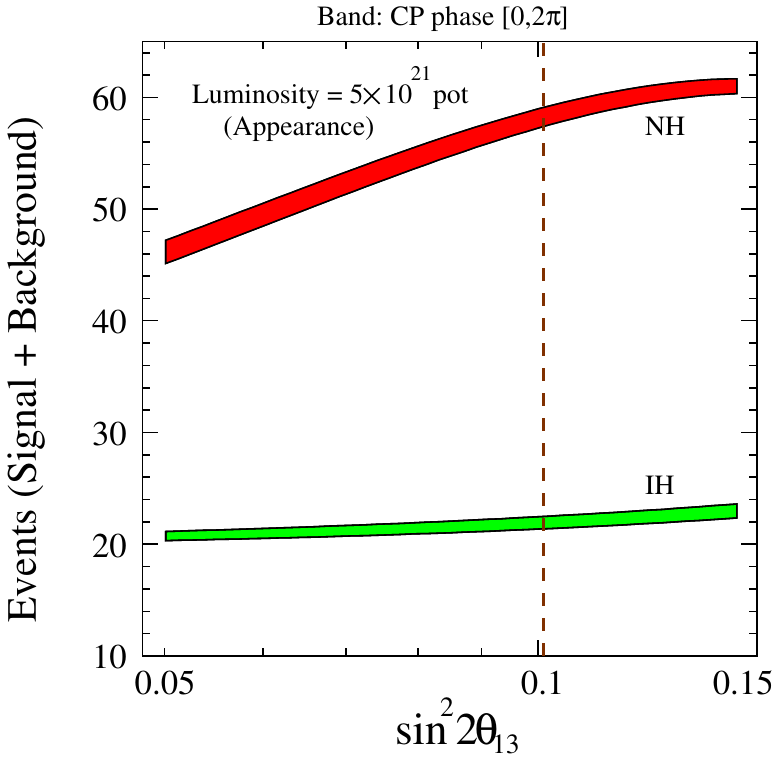}
\mycaption{\label{fig:event} The expected number of appearance events (signal + background) as a function of $\stch$ for the 
CERN-Kamioka baseline. The result is shown for a total luminosity of $5 \times 10^{21}$ pot. The band portrays the 
impact of unknown $\dcp$. The vertical dashed brown line shows the present best-fit value of $\stch$ of 0.101.}
\end{figure}

In Fig.~\ref{fig:event}, we show the expected number of events (signal + background) for the CERN-Kamioka baseline in the appearance 
channel as a function of $\stch$. Here we vary $\dcp$ in its allowed range of 0-2$\pi$ and the band reflects its impact. For $\stch$, we take the
$2\,\sigma$ allowed range as suggested in~\cite{Tortola:2012te}. We present the results for both the choices of the hierarchies taking a total
luminosity of $5 \times 10^{21}$ pot. The difference in the number of events for NH and IH is quite large even for the value of $\stch$ as small as 0.05.
The lower part of the red band suggests that for NH, the number of events increases from 45 to 60 while we increase $\stch$ from 0.05 to 0.14.

\subsection{Measurement of the Mass Ordering}
\label{sec:mo}

\begin{figure}[tp]
\includegraphics[width=0.6\textwidth]{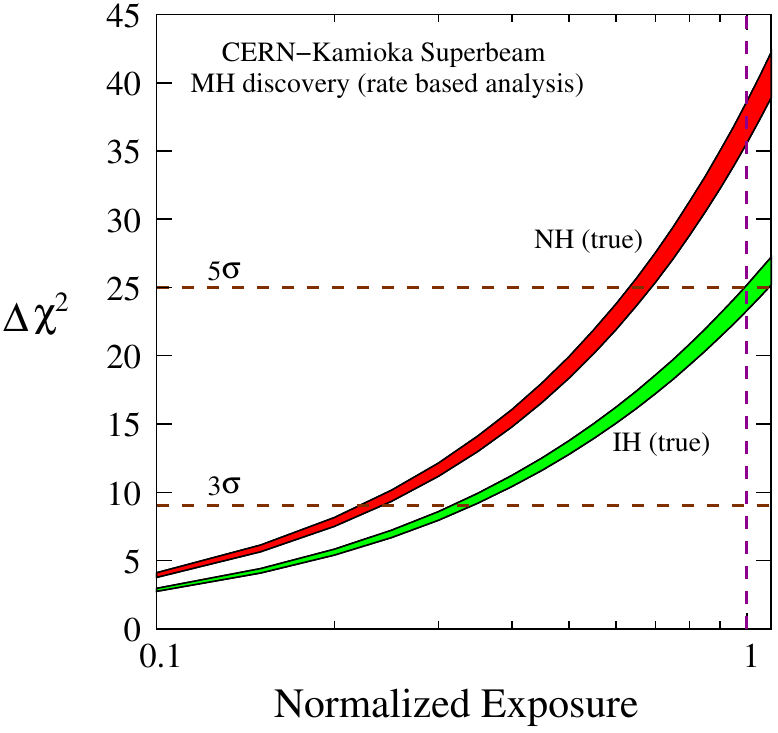}
\mycaption{\label{fig:surv_app} $\Delta\chi^{2}$ for the MH discovery as a function of the normalized exposure. 
Here, 1 in x-axis corresponds to $5\times10^{21}$ pot. The band portrays the effect of the unknown $\dcpt$. 
We have assumed $\stcht=0.101$ and for all other oscillation parameters, the central values are taken from
Table~\ref{tab:bench}. }
\end{figure}

Here we explore the capability of the set-up described in Sec.~\ref{sec:setup} to make a measurement of the 
true neutrino MH. A `discovery' of the MH is defined as the ability to exclude any degenerate solution for the 
wrong (fit) hierarchy at $5\,\sigma$ confidence level. For our sensitivity calculation, we include a 5\% systematic 
on the total number of signal events and a (uncorrelated) 5\% systematic on the total number of background events. 
They are included using the pull method as described in {\it e.g.} reference~\cite{Huber:2002mx,Fogli:2002au}. 
We perform the usual $\chi^2$ analysis using a Poissonian likelihood function adding the information coming from 
$\nue$ appearance and $\numu$ disappearance channel.

The total $\Delta \chi^2$ obtained for the wrong hierarchy hypothesis is shown in Fig.~\ref{fig:surv_app} as a function of 
the exposure normalized to our reference exposure of $5\times10^{21}$ pot. The two bands correspond to the true NH and 
IH cases respectively, the width of the band in each case corresponds to the variation of the true value of the phase $\dcp \in [0,2\pi]$. 
Fig.~\ref{fig:surv_app} shows that to measure the neutrino mass ordering at $5\,\sigma$ C.L. irrespective of the choice of 
true hierarchy and the value of $\dcpt$, we need an integrated luminosity of $5.4\times10^{21}$ pot.
To achieve the same at $3\,\sigma$ C.L., a total luminosity of $1.73\times10^{21}$ pot is required.

\begin{figure}[tp]
\includegraphics[width=0.6\textwidth]{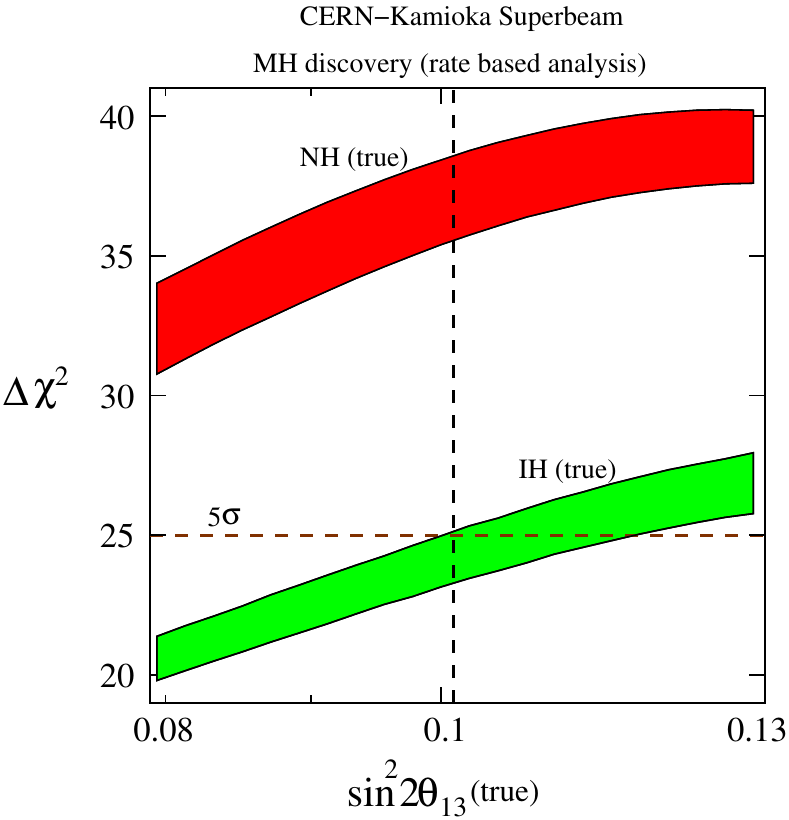}
\mycaption{\label{fig:sens-th13-vary} $\Delta\chi^{2}$ for the MH discovery as a function of $\stcht$.
Here we consider a total exposure of $5\times10^{21}$ pot. The band reflects the effect of the unknown $\dcpt$.
The vertical dashed black line shows the present best-fit value of $\stcht$.}
\end{figure}

In Fig.~\ref{fig:sens-th13-vary}, we address the issue that how the sensitivity of the proposed set-up towards MH discovery will change if we vary
$\stcht$ in its allowed $2\,\sigma$ range\footnote{We have used the same error on $\tet$ while changing its true value. This might not be the case
in real scenario.}. Here we consider our reference exposure of $5\times10^{21}$ pot. It is quite evident that with this 
exposure for true NH, this set-up can exclude IH well above $5\,\sigma$ C.L. for any values of $\dcpt$ in the allowed $2\,\sigma$ range of $\stcht$. 
We can accomplish the same at $4.4\,\sigma$ C.L. for true IH.

Here we would like to point out that there is probably margin for optimizing the signal-to-noise ratio further. For example, we have checked that 
increasing the threshold energy from $0.5\,\mathrm{GeV}$ to $4\,\mathrm{GeV}$, improves the sensitivity. Considering NH as true hierarchy
we need $2.9\times10^{21}$ pot ($3.4\times10^{21}$ pot) in the case of $4\,\mathrm{GeV}$ ($0.5\,\mathrm{GeV}$) threshold energy to exclude IH at
$5\,\sigma$ C.L. irrespective of the choice of $\dcpt$. However, a detailed Monte Carlo study is needed to perform this optimization in reality. 

Other baselines could be realistically considered such as Fermilab-Kamioka ($9160\,\mathrm{km}$) assuming the same superbeam would be 
build at Fermilab. We find slightly worse but very similar results for such longer baseline, everything else being the same. 
A more significant improvement is found for the shorter CERN-INO baseline ($7360\,\mathrm{km}$). In this case, however the detector would 
be an iron calorimeter, where it is more challenging to see electrons in the few $\mathrm{GeV}$ range. Reducing the thickness of iron
plates in the ICAL@INO detector similar to the MINOS detector may create an opportunity to observe the electrons. A detailed detector simulation 
study would be needed to estimate the efficiencies and backgrounds in that case. 

\section{Concluding Remarks}
\label{sec:conclusion}

In summary, we have explored a new strategy to determine the neutrino MH that employs a superbeam, as those being proposed at CERN, focused 
towards the {\it existing}  and well-understood Super-Kamiokande detector, $8870\,\mathrm{km}$ away. We have shown that the neutrino beam 
resulting from a proton source with a total exposure of 5.4$\times10^{21}$ pot can reveal the neutrino MH at $5\,\sigma$ irrespective of the true hierarchy 
and CP phase, via a simple counting of $\nu_e$ events at the far location. We believe there is margin for optimization both in the neutrino beam energy 
and in the detector signal-to-noise ratio. The proposed measurement of the neutrino MH would probably come earlier, and therefore fit nicely as a first step 
in the long-term plan to measure the more subtle effect of the CP violating phase. The latter requires a more ambitious program involving, for example, 
a superbeam running in both polarities, together with a significant upgrade in the detector technology, {\it e.g.} a factor of $20$ increase in the size of 
Super-Kamiokande, as proposed in  the T2HK project~\cite{Abe:2011ts}, or a more challenging detector technology such as 
${\mathcal O}(20)$ $\mathrm{kt}$~Liquid Argon detector \cite{Agarwalla:2011hh}.

\acknowledgments

We would like to thank Andrea Longhin, Carlos Pe\~na-Garay and Andr\'e Rubbia for useful discussions. 
This work was partially supported by the Spanish Ministry for Education and Science projects  
FPA2007-60323, FPA2011-29678; the Consolider-Ingenio CUP (CSD2008-00037) and CPAN (CSC2007-00042); 
the Generalitat Valenciana (PROMETEO/2009/116); the European projects EURONU (CE212372) and LAGUNA (Project Number 212343).

\bibliographystyle{apsrev} 
\bibliography{references}

\end{document}